\documentclass[aps,prd,twocolumn,superscriptaddress,showpacs,nofootinbib]{revtex4-2}

\usepackage{epsfig}
\usepackage{graphicx}
\usepackage{dcolumn}
\usepackage{bm}
\usepackage{multirow}
\usepackage{array}
\usepackage{slashed}
\usepackage{amsmath}
\usepackage[titletoc]{appendix}
\usepackage{color}
\usepackage{url}
\usepackage{hyperref}
\usepackage{soul}
\usepackage{xcolor}

\begin{document}

\preprint{APS/123-QED}

\title{\textbf{Exploring the properties of newborn pulsars with high-energy neutrinos}}%

\author{Zi-Zhuo Xiao}
\affiliation{School of Mathematics and Physics, China University of Geosciences, Wuhan 430074, China}

\author{Gang Guo}
\email{Corresponding author: guogang@cug.edu.cn} 
\affiliation{School of Mathematics and Physics, China University of Geosciences, Wuhan 430074, China}
\affiliation{Shenzhen Research Institute of China University of Geosciences, Shenzhen 518057, China}

\author{Tuohuniyazi Tuniyazi}
\affiliation{School of Mathematics and Physics, China University of Geosciences, Wuhan 430074, China}

\date{\today}

\begin{abstract}
Newborn pulsars resulting from core-collapse supernovae (CCSNe) are promising sources of high-energy (HE) cosmic rays and neutrinos. In this work, we focus on HE neutrinos generated by interactions between protons accelerated in relativistic pulsar winds and SN ejecta. Using a binned likelihood analysis, we evaluate the detection prospects of these neutrinos with IceCube and explore their potential for probing the initial spin period ($P$) and magnetic field strength ($B$) of newborn pulsars. Noticing a degeneracy in neutrino signal with $B/P^2$ across a broad parameter space, we find that HE neutrinos from a galactic newborn pulsar ($d=10$ kpc) with $(B/10^{12}~{\rm G})(P/{\rm ms})^{-2} \gtrsim 0.003$ can be detected at $\gtrsim 3\sigma$ significance. Even in the absence of a nearby pulsar, the diffuse flux from the cosmologic population of pulsars could be probed by next-generation detectors like IceCube-Gen2 and GRAND200k. For galactic pulsars with $(B/10^{12}~{\rm G})(P/{\rm ms})^{-2} \gtrsim 0.08$, the combination $B/P^2$ can be measured with an accuracy of better than 20\%  
at the $3\sigma$ confidence level.
Additionally, we show that for pulsars with detectable HE neutrino signals, the emission can be clearly distinguished from neutrinos produced by interactions between SN ejecta and the circumstellar medium, due to their distinct temporal and spectral features.
\end{abstract}

\maketitle


\section{introduction}

The discovery of high-energy (HE) astrophysical neutrinos over the past decades has ushered in a new era of astronomy, offering crucial insights into cosmic ray (CR) acceleration and the nature of extreme astrophysical environments \cite{Aartsen_2013,Evidence_for_HE_2013}. Notably, the established associations of IceCube HE neutrino events with active galactic nuclei (AGN), such as TXS 0506+056 \cite{TXS:2018}  and NGC 1068 \cite{NGC:2022}, have confirmed AGNs as promising sources of HE CRs and neutrinos, while also constraining the related radiation models \cite{Keivani:2018rnh,Gao:2018mnu,Rodrigues:2018tku,Cerruti:2018tmc,Sahakyan:2018voh,Reimer:2018vvw,Padovani:2019xcv,Zhang:2019htg,Petropoulou:2019zqp,Xue:2019txw,Murase:2019vdl,Inoue:2019yfs,Fang:2023vdg,Padovani:2024ibi,Das:2024vug,Yasuda:2024fvc}. There is also growing evidence linking a handful of HE neutrino events to known tidal disruption events (TDEs) \cite{Stein:2020xhk,Reusch:2021ztx,vanVelzen:2021zsm,Yuan:2024foi,Li:2024qcp}. Despite these advances, the origins of the majority of observed HE neutrino events at IceCube, as well as the most energetic event recently observed by KM3NeT \cite{km3net1,km3net2}, remain unknown. Their diverse arrival times and directions suggest they likely originate from numerous distant, individual sources yet to be identified.

Among the candidate sources, core-collapse supernovae (CCSNe) represent a distinct and compelling possibility. These events, which mark the catastrophic end of massive stars,
release nearly all of their gravitational binding energy ($\sim 3\times 10^{53}$~erg) in the form of low-energy (LE) neutrinos in the 10–100 MeV range \cite{Janka:2006fh,Janka:2012wk,Burrows:2013,Janka:2017vlw,Fischer:2023ebq}, as confirmed by the detection of neutrinos from the nearby SN1987A \cite{SK:SN1987A,IBM:SN1987A}. In addition to these LE neutrinos, CCSNe could also produce non-thermal HE neutrinos through hadronic processes involving accelerated protons or nuclei during or after the explosion \cite{Tamborra:2018upn,Tamborra:2024fcd}. Intriguingly, Ref. \cite{Oyama:2021caa} has suggested a potential correlation between SN 1987A and HE neutrino events, raising the possibility that CCSNe could serve as sources of detectable HE neutrinos. Unlike AGNs or TDEs, which are typically distant and produce isolated events, CCSNe can occur within or near our Galaxy \cite{Adams:2013ana,Rozwadowska:2020nab}, making them especially valuable targets for high-statistics, time-resolved neutrino observations. 
Combined with electromagnetic observations and gravitational wave signals,
the simultaneous detection of both LE and HE neutrinos from a future galactic CCSN \cite{Guetta:2023mls,Ashida:2024nck,Zegarelli:2024ivy,Guo:2023sbt} would provide an unprecedented multi-messenger window into stellar collapse, particle acceleration, and extreme astrophysical radiation processes.

Several mechanisms have been proposed for HE neutrino production in CCSNe above the GeV scale. One well-established possibility involves proton acceleration at collisionless shocks formed as the SN ejecta interact with the surrounding circumstellar medium (CSM) \cite{Chevalier1994,Chevalier:2001at,Chevalier:2016hzo}. These energetic protons can undergo $pp$ collisions with the CSM, generating charged pions and kaons that decay into HE neutrinos \cite{Murase:2010cu,Murase:2017pfe,Murase:2018okz,Kheirandish:2022eox,Sarmah:2022vra,Valtonen-Mattila:2022nej,Murase:2023chr,Kimura:2024lvt}.
Another scenario is the choked jet model, where relativistic jets form but fail to escape the stellar envelope, resulting in efficient HE neutrino production without accompanying gamma-rays \cite{Meszaros2001,Razzaque_2003,Razzaque2004,Ando:2005xi,RAZZAQUE_2005,Horiuchi:2007xi,Bartos:2012sg,Murase_2013,Xiao2014,Varela2015,Tamborra2015,Senno2015,Senno:2017vtd,Denton:2017jwk,Denton:2018tdj,He:2018lwb,Chang:2022hqj,Guarini:2022hry,Guo:2022zyl}. 

Beyond these shock-related mechanisms,
a significant fraction of CCSNe lead to the formation of young pulsars—rapidly rotating neutron stars (NSs) with
strong magnetic fields, which have been suggested as potential sources of HE CRs and neutrinos \cite{Blasi_2000,Zhang:2002xv,Fang_2012,Fang_2013}. Particle acceleration can occur within the pulsar magnetosphere, particularly in the polar cap, slot gap, and outer gap regions where magnetic energy is converted into kinetic energy of charged particles \cite{Ruderman1975,Arons1983,Cheng1986}. Above the magnetosphere, the relativistic pulsar wind undergoes acceleration at the termination shock via Fermi processes and magnetic reconnection in the striped wind \cite{Lemoine_2015}. In pulsar wind nebulae (PWNe), additional acceleration may occur as pulsar winds interacting with the SN remnants \cite{Gaensler:2006ua}. These accelerated protons or nuclei can then interact with the radiation fields or baryons inside the source, leading to HE neutrino production \cite{Nagataki:2003kq,Murase_2009,Fang_2013,Fang:2014qva}.

While the LE neutrino signal from CCSNe has been extensively studied \cite{Mirizzi:2015eza,Horiuchi:2018ofe}, HE neutrinos have received comparatively less attention.
In particular, the potential to use HE neutrinos as probes of CCSNe or their compact remnants, such as pulsars, remains largely unexplored. This question is especially timely given the possibility of a future galactic CCSN. In this work, 
we focus on HE neutrino production from newly formed pulsars, evaluate their observability with current and next-generation neutrino telescopes, and examine how these signals could constrain key pulsar properties such as magnetic field strength and initial spin period.
Specifically, we adopt a minimal pulsar scenario for HE neutrino production and carry out a likelihood analysis to evaluate the detection prospect of the neutrino signals and their observational implications.

The structure of this paper is as follows.  In Sec.~\ref{sec:pulsar-model}, we review HE neutrino production within the minimal pulsar scenario and present the resulting fluences. Sec.~\ref{sec:signals} focus on the detection prospects for HE neutrinos from both a nearby newborn pulsar and the diffuse fluxes from the cosmic pulsar population. In Sec.~\ref{sec:pulsar properties}, we explore the potential to probe a pulsar's magnetic field strength and initial spin period using the HE neutrino signals. Sec.~\ref{sec:pulsar-or-not} compares the pulsar scenario with the ejecta-CSM interaction model, assessing the feasibility of distinguishing between these two mechanisms through their distinct neutrino signatures. Finally, we summarize and conclude our findings in Sec.~\ref{sec:summary}.

\section{HE Neutrino from pulsars}
\label{sec:pulsar-model}

In this section, we review the production of HE neutrinos within the minimal pulsar model \cite{Fang:2013vla,Carpio:2020wzg}. Pulsars lose rotational energy as relativistic winds are launched from their surfaces. Assuming
the energy loss is governed by electromagnetic dipole radiation, the spindown luminosity at time $t$ is given by
\begin{align}
L_{\rm sd}(t) \approx& {5 B^2 R_{\rm NS}^6 \Omega^4 \over 12c^3} \Big(1+{t\over t_{\rm sd}}\Big)^{-2} \nonumber \\
\approx& 1.5 \times 10^{50}~{\rm erg~s^{-1}}~B^2_{15}R_{\rm NS, 6}^6\Omega_{4}^4 \nonumber \\
& \times \Big(1+{t\over t_{\rm sd}}\Big)^{-2}, \label{eq:Lsd}
\end{align}
where $B$ is the surface magnetic field strength, $R_{\rm NS}$ is the NS radius, $\Omega=2\pi/P$ with $P$ being the initial spin period, and $c$ is the speed of light. The characteristic spindown time $t_{\rm sd}$ is \cite{Murase_2016}
\begin{align}
	t_\text{sd} ={6I c^3 \over 5\Omega_\text{i}^2B_\text{NS}^2R_\text{NS}^6}
   \simeq 10^{2.5}~\text{s}~ I_{45}B_{ 15}^{-2}R_{\text{NS}, 6}^{-6}\Omega_4^{-2},
\label{eq:tsd}
\end{align}
where $I$ is the NS moment of inertia, and we adopt $R_{\rm NS}=10^{6}$~cm and $I=10^{45}$~g~cm$^{2}$ throughout this work. Unless otherwise specified, we use the notation $A_x \equiv A/10^x$, with $A$ expressed in cgs units. Exceptions include the energies of HE protons and neutrinos, which are expressed in GeV, and the spin period $P$, which is given in milliseconds (ms).

Charged particles can be accelerated by the electric field induced by the rotating magnetic field in the pulsar winds. For simplicity, we assume the accelerated nuclei are purely protons. Provided that these protons experience an effective fraction $\eta$ of the potential gap, protons at time $t$ can be accelerated to the energy \cite{Murase_2009,Arons_2003}
\begin{eqnarray}\nonumber
	E_p(t)&  = & \frac{\eta eBR_\text{NS}^3}{2c^2}\Omega_\text{i}^2\\\nonumber 
	& \simeq & 1.7\times 10^{12}~\text{GeV}\,  \eta_{-1}B_{15}R_{\text{NS},6}^{3}\Omega_4^{2}\\
	& & \times\Big(1+{t\over t_{\rm sd}}\Big)^{-1}.
\label{eq:Ept}
\end{eqnarray}

Despite the dominance of $e^\pm$ pairs in pulsar winds, the ion injection rates could be comparable to the Goldreich-Julian rate \cite{Goldreich:1969sb} if a significant fraction of the energy is carried by ions. Assuming this rate applies to protons, their injection rate at $t$ is
\begin{eqnarray} 
\dot N_p(t) & = & \frac{BR_{\text{NS}}^3\Omega^2}{ec}\Big(1+{t\over t_{\rm sd}}\Big)^{-1} \nonumber \\
  & \simeq & 7\times 10^{39}~{\rm s}^{-1}~B_{15}R_{\rm NS,6}^3\Omega_4^{2}
 \Big(1+{t\over t_{\rm sd}}\Big)^{-1}.
\label{eq:Npdot}
\end{eqnarray}
From Eqs.~\eqref{eq:Ept} and \eqref{eq:Npdot}, it follows that $E_p(t)\dot N_p(t)\sim 0.1 \eta_{0.1}L_{\rm sd}(t)$, implying that accelerated protons carry approximately 10\% of the total spindown luminosity.  

As these protons propagate through the expanding SN ejecta, they undergo hadronic interactions via $pp$ processes, producing charged mesons (e.g., $\pi^\pm$ and $K^\pm$) that subsequently decay into HE neutrinos. At early times ($t\lesssim 10^{2-3}$ years), the ejecta radius grows linearly as $r_{\rm ej}(t)=c\beta_{\rm ej} t$, where $\beta_{\rm ej}$ is the initial ejecta velocity in units of the speed of light. Assuming spherical symmetry, the proton density in the ejecta is $n_{p}(t) = 3M_\text{ej}/[4\pi r_\text{ej}^3(t) m_p]$, with $M_{\rm ej}$ the total ejecta mass and $m_p$ the proton mass. The optical depth for $pp$ interactions for an accelerated proton with energy $E_p(t)$ can be estimated as 
\begin{align}
f_{pp}(E_p(t)) &\sim \min[1,~n_p(t) r_{\rm ej}(t) \sigma_{pp}] \nonumber \\
& \approx \min[1,~3\times 10^{5} M_{{\rm ej},1}\beta_{{\rm ej},-1}^{-2}t_4^{-2} \sigma_{pp,-25}],
\label{eq:fpp}
\end{align}
where $M_{\rm ej, 1} \equiv {M_{\rm ej}\over 10 M_{\odot}}$, and $\sigma_{pp}$ is the total cross section for $pp$ collisions.

The emitted neutrino spectrum during a given time window $[t_{\rm s}, t_{\rm e}]$ after explosion is numerically computed as below. We first divide the time window into small intervals. During each time step, a total number of $\dot N_p(t_i) \delta t_i$ protons with energy $E_p(t_i)$ are emitted [see Eqs.~\eqref{eq:Ept} and \eqref{eq:Npdot}]. Considering the optical depth for $pp$ interaction given in Eq.~\eqref{eq:fpp}, the spectra of charged mesons generated at the time step can be expressed as 
\begin{align}
{dN_m^{(i)} \over dE_m} =        
\dot N_p(t_i) \delta t_i f_{pp}(E_p(t_i)) Y_{pp \to m}(E_p(t_i), E_m),\label{eq:spec_meson}
\end{align}
where $Y_{pp \to m}$ represents the averaged meson spectrum produced per $pp$ reaction, and $m$ stands for $\pi^\pm$ or $K^\pm$. In our study, we use PYTHIA 8.3 \cite{Bierlich:2022pfr} to obtain the meson yields. In the relevant energy  range, $\pi^-$ and $\pi^+$ yields are approximately equal, as are those for $K^\pm$.

We then simulate the decays of $\pi^\pm$ and $K^\pm$, explicitly including $\pi^- \to \mu^- + \bar\nu_\mu$, $K^- \to \mu^- + \bar\nu_\mu$, $\mu^- \to e^- + \nu_\mu + \bar\nu_e$, along with the corresponding decays for $\pi^+$, $K^+$, and $\mu^+$. For $K^\pm$ decays into $\mu^\pm$, a branching ratio of $f_K \approx 0.64$ is taken into account. To model the three-body muon decay, we adopt the analytical expressions provided in Ref.~\cite{Lipari:2007su}. Charged mesons could loss energy through interactions with the SN ejecta before decaying, leading to a suppression of the HE neutrino yield. This suppression for each meson type is quantified by
a factor $f_m$ (with $m=\pi^\pm, K^\pm, \mu^\pm$) \cite{Carpio:2020wzg}:
\begin{align}
f_m(E_m) = 1-\exp(-t_{mp}/t_{m,\rm dec}), \label{eq:f_cool}
\end{align}
where $t_{m,\rm dec} =(E_m/m_m)\tau_m$ with $m_m$ and $\tau_m$ denoting the mass and lifetime of charged meson $m$, respectively. The cooling time due to meson-proton scattering is estimated as $t_{mp} \approx (\kappa_{m p}\sigma_{m p} n_p c)^{-1}$, where $\kappa_{m p}$ represents the average inelasticity and $\sigma_{m p}$ is the total cross section. We take the energy-dependent cross sections $\sigma_{\pi^\pm p}$ and $\sigma_{K^\pm p}$ from PYTHIA 8.3 \cite{Bierlich:2022pfr} and assume $\kappa_{\pi^\pm p}=\kappa_{K^\pm p}=0.8$. To account for muon cooling, which is dominated by bremsstrahlung,  pair production, and photonuclear interactions, we adopt $\kappa_{\mu p}\sigma_{\mu p }=10^{-29}$~cm$^2$ \cite{Carpio:2020wzg}.

\begin{figure*}[htbp]
\centering   \includegraphics[width=0.49\textwidth]{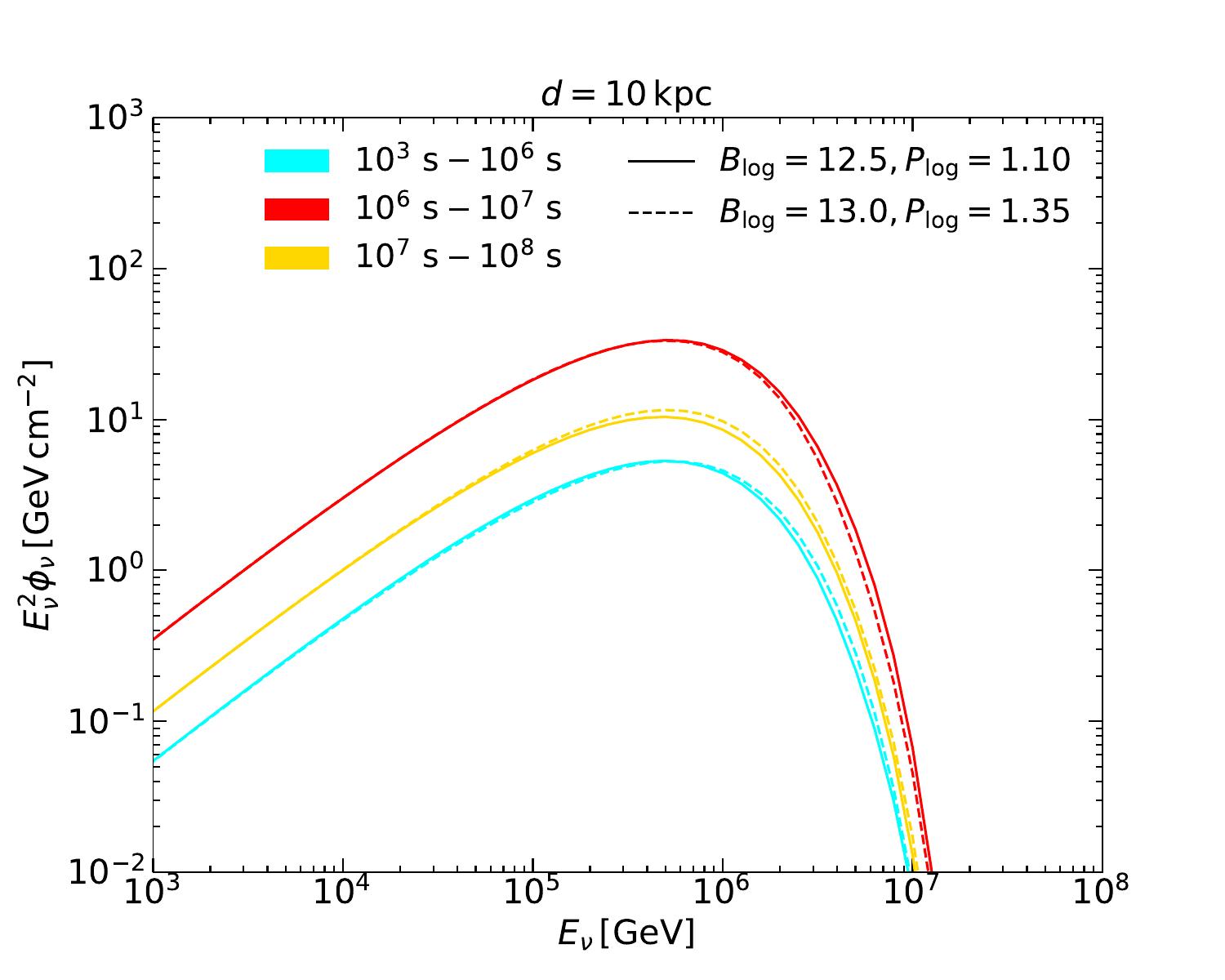}
\includegraphics[width=0.49\textwidth]{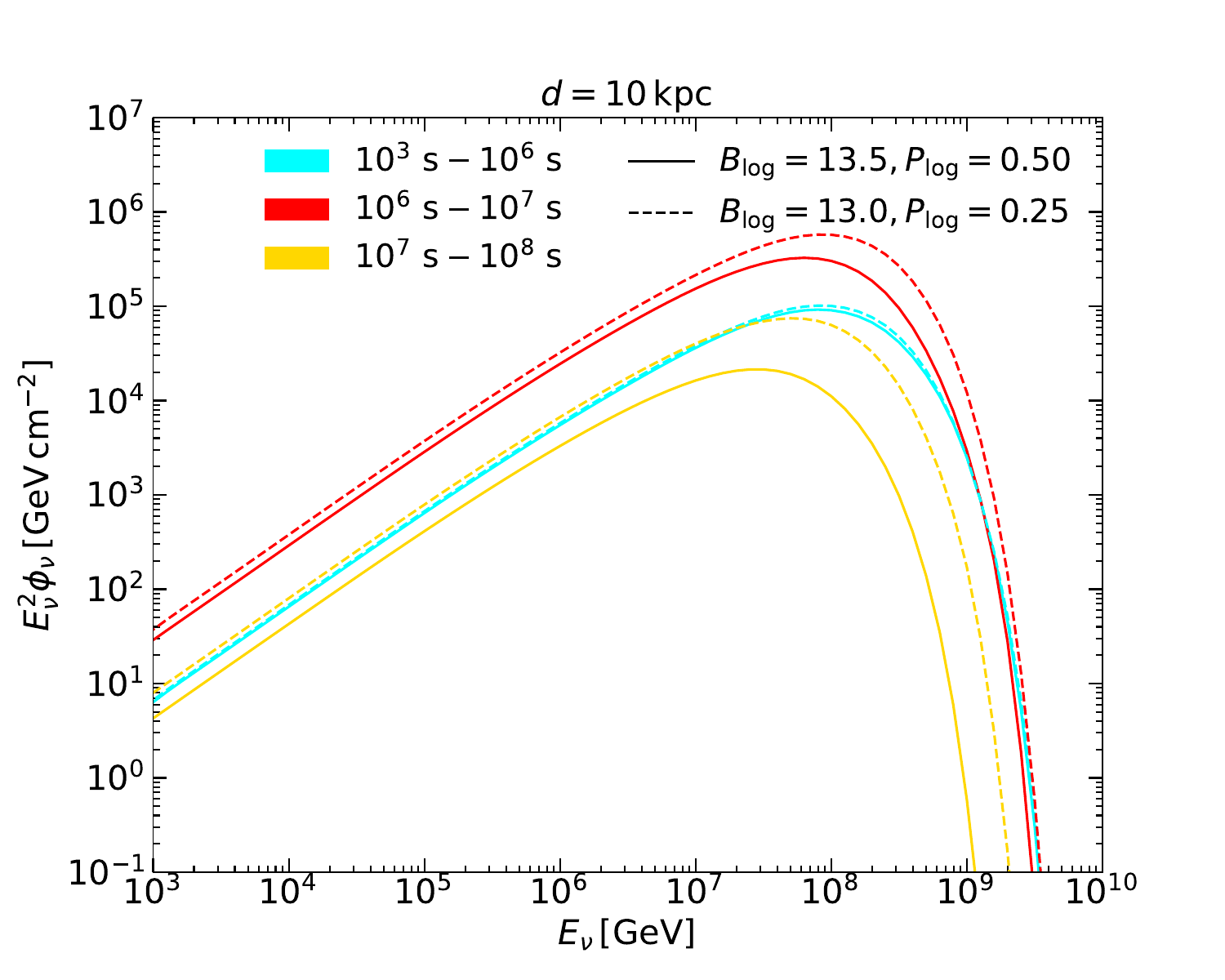}
\caption{Neutrino fluences from newborn pulsars at 10 kpc with different values of $B$ and $P$ integrated over different time windows. Left panel: slowly rotating pulsars with $(B_{\rm log}, P_{\rm log})\equiv (\log_{10} (B/{\rm G}), \log_{10} (P/{\rm ms}))=(12.5, 1.1)$ and (13.0, 1.35). Right panel: rapidly rotating pulsars with $(B_{\rm log}, P_{\rm log})=(13.0, 0.25)$ and (13.5, 0.5).}
\label{fig:fluence}
\end{figure*}

Based on the spectra of $\pi^\pm$ and $K^\pm$ provided in Eq.~\eqref{eq:spec_meson}, we
simulate their decays to calculate the resulting HE neutrino yields.
By summing contributions from all time steps between $t_{\rm s}$ and $t_{\rm e}$, we obtain the total emitted neutrino spectrum (in unit of $\rm GeV^{-1}$) during the time window, denoted as $dN_\nu/dE_\nu(E_\nu, B, P)$.
The corresponding all-flavor neutrino fluence from a single pulsar, in unit of $\rm GeV^{-1}~cm^{-2}$,  is then given by
\begin{align}
F_\nu(E_\nu, B, P) = {1 \over 4\pi d^2} {dN_\nu \over dE_\nu}(E_\nu, B, P),
\end{align}
where $d$ denoted the distance to the pulsar.

Figure~\ref{fig:fluence} shows the resulting neutrino fluences (multiplied by $E_\nu^2$) from a newborn pulsar located at 10 kpc, evaluated over three distinct time windows: $[10^3, 10^6]$~s, $[10^6, 10^7]$~s, and $[10^7, 10^8]$~s. In the left panel, we consider $(B_{\rm log}, P_{\rm log}) \equiv (\log_{10}(B/{\rm G}), \log_{10}(P/{\rm ms}))=(12.5, 1.1)$ and (13.0, 1.35), while the right panel shows the results for (13.0, 0.25) and (13.5, 0.5). The spindown times for the four cases are approximately $1.2\times 10^{10}$~s, $4\times 10^9$~s, $8\times 10^6$~s, and $2.5\times 10^7$~s, respectively [Eq.~\eqref{eq:tsd}]. For the above chosen parameter sets, most HE neutrinos are produced during the window of $10^6$--$10^7$~s. We have also verified that this conclusion remains valid across most of the parameter space explored in this work ($10^{12}<B<10^{14}$~G and $1<P<100$~ms).
At early times ($10^3$--$10^6$ s), the neutrino fluence is primarily constrained by the number of accelerated protons and, to a lesser extent, by the cooling of charged mesons in the dense ejecta [Eq.~\eqref{eq:f_cool}]. As the ejecta expands, the optical depth for $pp$ reaction, $f_{pp}$, decreases [Eq.~\eqref{eq:fpp}], resulting in a lower neutrino yield at very late times.

The energies of neutrinos from newborn pulsars range from $10^{4}$--$10^{10}$~GeV.
Stronger magnetic fields or faster rotation increase both the proton injection rate and their maximum energy [Eqs.~\eqref{eq:Ept} and \eqref{eq:Npdot}], thereby leading to a larger HE neutrino fluence that peaks at a higher $E_\nu$. Notably, aside from the additional dependencies due to $t_{\rm sd}$, both the proton energy $E_p(t)$ and its injection rate $\dot N_p(t)$ scale with the combination $B/P^2$. As a consequence, the neutrino fluence exhibits a degeneracy between $B$ and $P$ in certain regions of the parameter space. At $t \ll t_{\rm sd}$, different parameter combinations sharing the same $B/P^2$ produce nearly identical neutrino fluences (see the left panel of Fig.~\ref{fig:fluence}). Such degeneracy breaks down when $t$ becomes comparable or exceeds $t_{\rm sd}$ and the factor $(1+{t \over t_{\rm sd}})$ begins to affect the fluences. As shown in the right panel,  
the fluences for two cases with the same $B/P^2$ differ noticeably in the time windows of $10^6$--$10^7$~s and $10^7$--$10^8$~s.

\section{detection prospect of HE neutrinos from pulsars}
\label{sec:signals}

In this section, we study the detection prospects of HE neutrinos originating from newborn pulsars. 
We begin by analyzing the expected neutrino signals from individual nearby sources, particularly those to occur within our Galaxy. Owing to their proximity and the distinct temporal and spectral characteristics of the resulting neutrino emissions, these galactic pulsars represent the most promising candidates for detection.
In addition, we evaluate the cumulative contribution of the entire population of newborn pulsars to the diffuse neutrino background. This diffuse component becomes especially relevant in scenarios where individual sources fail to produce detectable signals above background levels.

\subsection{neutrino signals from a nearby pulsar}

To assess the detectability of HE neutrinos from a nearby pulsar, we concentrate on the IceCube detector  \cite{IceCube:2016zyt}. To optimize the detection sensitivities, we employ upgoing tracks for studying HE neutrinos from the northern hemisphere, and starting cascades for those from the southern hemisphere. For a nearby newborn pulsar, we fix $\cos\theta=0.5$ and $-0.5$, corresponding to pulsars located in the southern and northern hemispheres, where $\theta$ is the zenith angle of the neutrino arrival direction.

\begin{figure}[htbp]
\centering    \includegraphics[width=0.49\textwidth]{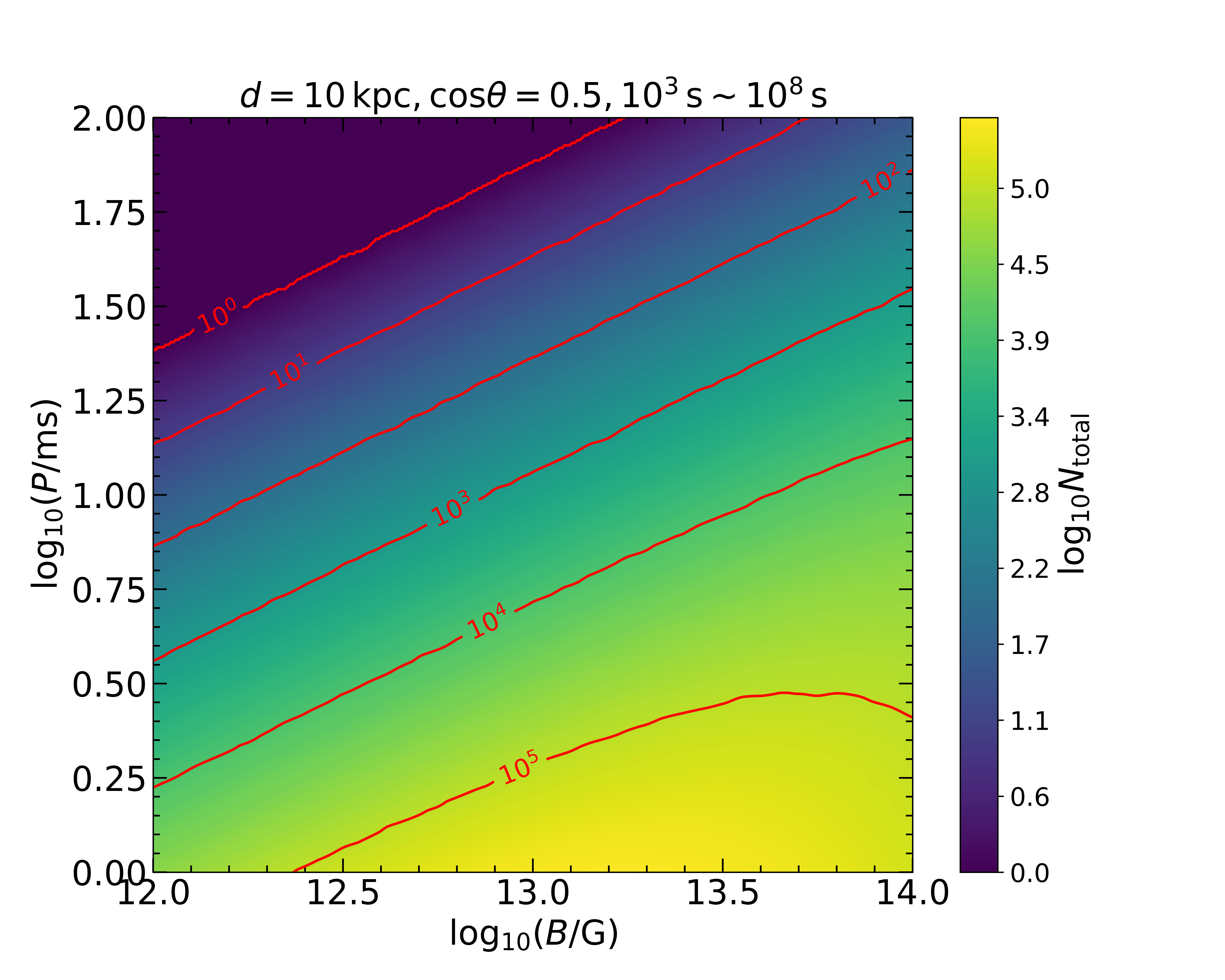}
\caption{Expected total number of neutrino-induced cascade events from a nearby newborn pulsar located at 10 kpc, for different values of magnetic field strength $B$ and initial spin period $P$. The event counts are integrated over the time window $[10^3, 10^8]$ s, with deposited energy above 100 GeV. The red lines indicate contours of $N_{\rm tot}$. Pulsars are assumed to located in the southern hemisphere with $\cos\theta=0.5$ and downgoing cascades are considered.}
\label{fig:Ntot}
\end{figure}

Upgoing tracks are primarily produced by the charged-current (CC) interactions of $\nu_\mu$ and $\bar\nu_\mu$, with a smaller constribution from the CC interactions of $\nu_\tau$ and $\bar\nu_\tau$. This channel offers large effective areas, making it suited for detecting astrophysical neutrinos from the northern hemisphere at IceCube.
However, for HE neutrinos from the southern hemisphere, atmospheric muons could contribute significantly to the background rate. To address this, we focus on
starting cascade events, where the neutrino interaction vertex is contained within the instrumented volume of the detector. These cascades mainly arise from neutral-current (NC) interactions of all neutrino flavors, along with CC interactions involving electron and tau neutrinos. Starting cascades provide several advantages, including accurate energy reconstruction and being largely free from contamination by atmospheric muons. This makes them particularly well-suited for identifying astrophysical neutrinos from the Southern hemisphere with IceCube \cite{IceCube:2014rwe,IceCube:2019lzm,IceCube:2021mbf,IceCube:2023ame}.
Note that the use of cascade events is crucial for studying HE neutrinos from galactic CCSNe or pulsars, as these events are expected to occur preferentially near the Galactic center, where the associated HE neutrinos are more likely to reach the IceCube detector from downgoing directions.

We employ a binned maximum likelihood analysis to evaluate the detection sensitivities. Our approach adopts a two-dimensional binning scheme in reconstructed deposited energy $E_{\rm d}$ and arrival time $t$ relative to the SN explosion. For HE neutrinos from a nearby pulsar, the expected number of tracks or cascade events in time bin 
$i$ (bounded by $t_i^{\rm lower}<t\le t_i^{\rm upper}$) and energy bin $j$ is given by
\begin{align}
\mu_{ij} = \sum_{f, C} \int_{t_i^{\rm lower}}^{t_i^{\rm upper}} dt \int dE_\nu \phi_f(E_\nu, t) A_{f,C,j}(E_\nu).  \label{eq:mu_ij}
\end{align}
Here, $f \in \{\nu_{e,\mu,\tau}, \bar\nu_{e,\mu,\tau}\}$ represents the neutrino flavor, $C \in \{\rm CC, NC\}$ denotes the interaction type, $\phi_f(E_\nu, t)$ (in unit of $\rm GeV^{-1}~cm^{-2}~s^{-1}$) is the time-dependent neutrino flux at Earth of flavor $f$, 
and $A_{f,C,j}(E_\nu)$ is the IceCube effective area for a neutrino of flavor $f$ and energy $E_\nu$ to produce an event with $E_{\rm d}$ in the $j$-th bin via interaction type $C$.
In our analysis we do not include angular binning for neutrino events, as the arrival direction of HE neutrinos can be well reconstructed--typically $\sim 1^\circ$ for track events and $\sim 10^\circ$ for cascades.

For upgoing tracks, we adopt the effective areas of IceCube from~\cite{IceCube:2015qii,track-effetive-area}. For starting cascade events induced by HE neutrinos from the southern hemisphere, we use the effective areas for the socalled medium-energy-starting events (MESEs) from~\cite{IceCube:2014rwe,cascade-effetive-area}, which are comparable to those used in an updated study \cite{IceCube:2019lzm}. It is important to note that recent developments in deep learning–based event reconstruction have demonstrated the ability to enhance the effective area for cascade events by a factor of 3–10 at TeV–PeV energies \cite{IceCube:2021mbf,IceCube:2023ame}. To incorporate such improvements, we scale the tabulated effective areas of \cite{IceCube:2014rwe} for cascade events by a factor of 3 in our study. In the binned likelihood analysis, we use the same energy bins for $E_{\rm d}$ as provided in the effective area tables, and 5 time bins equally spaced in logarithmic scale ranging from $10^3$ s to $10^8$ s.

Potential background sources include atmospheric neutrinos and muons. For both upgoing tracks and starting cascade events, atmospheric muons contribute significantly less than atmospheric neutrinos \cite{IceCube:2015qii,IceCube:2014rwe,IceCube:2019lzm,IceCube:2021mbf,IceCube:2023ame} and are therefore neglected in our analysis.
For starting cascades, while some atmospheric neutrinos can be vetoed through the identification of accompanying muons from the same cosmic ray air shower, we neglect this effect for the sake of a conservative estimate of detection sensitivity.
Diffuse astrophysical neutrinos represent another irreducible source of background. To estimate the expected number of background events in each bin, we apply Eq.~\eqref{eq:mu_ij} using fluxes of atmospheric obtained from the MCEq code \cite{MCEq} and diffuse astrophysical neutrinos from Refs. \cite{IceCube:2014rwe} and \cite{IceCube:2015qii} for starting cascade and upgoing tracks, respectively. 
We adopt a fixed solid angle of $\Delta \Omega = 0.1$ for starting cascades and 0.001 for tracks
as the search window,\footnote{We have verified that our results are insensitive to the choices of  $\Delta\Omega$, given the low background levels.} corresponding to a cone angle of $\sim 10^\circ$ and $\sim 1^\circ$ around the pulsar direction, respectively. These choices ensure effective inclusion of HE neutrinos from the source while suppressing the atmospheric backgrounds. The estimated background rate due to atmospheric and diffuse astrophysical neutrinos should be multiplied by the factor $\Delta\Omega$.

Figure~\ref{fig:Ntot} shows the expected total number of neutrino events, $N_{\rm tot}$, integrated over the time window [$10^3$, $10^8$]~s with deposited energy above $10^2$ GeV at IceCube, as a function of $B$ and $P$. For illustration, we consider a galactic newborn pulsar located at a distance of $d=10$ kpc and assume $\cos\theta=0.5$ (downgoing cascade events). As expected, stronger $B$ and/or shorter $P$ leads to higher neutrino event rates. The contours of $N_{\rm tot}=1$, 10, $10^2$, $10^3$, $10^4$, and $10^5$ in the 
$B_{\rm log}$--$P_{\rm log}$ plane
are denoted by red curves, reflecting a clear degeneracy along lines of constant $B/P^2$ in the region of weaker fields and slower rotation (e.g., $N_{\rm tot}=1$, 10, $10^2$, $10^3$, and $10^4$). This degeneracy becomes partially broken in the region of stronger $B$ and shorter $P$, as indicated by the $N_{\rm tot}=10^5$ contour.

\begin{figure*}[htbp]
\centering   \includegraphics[width=0.49\textwidth]{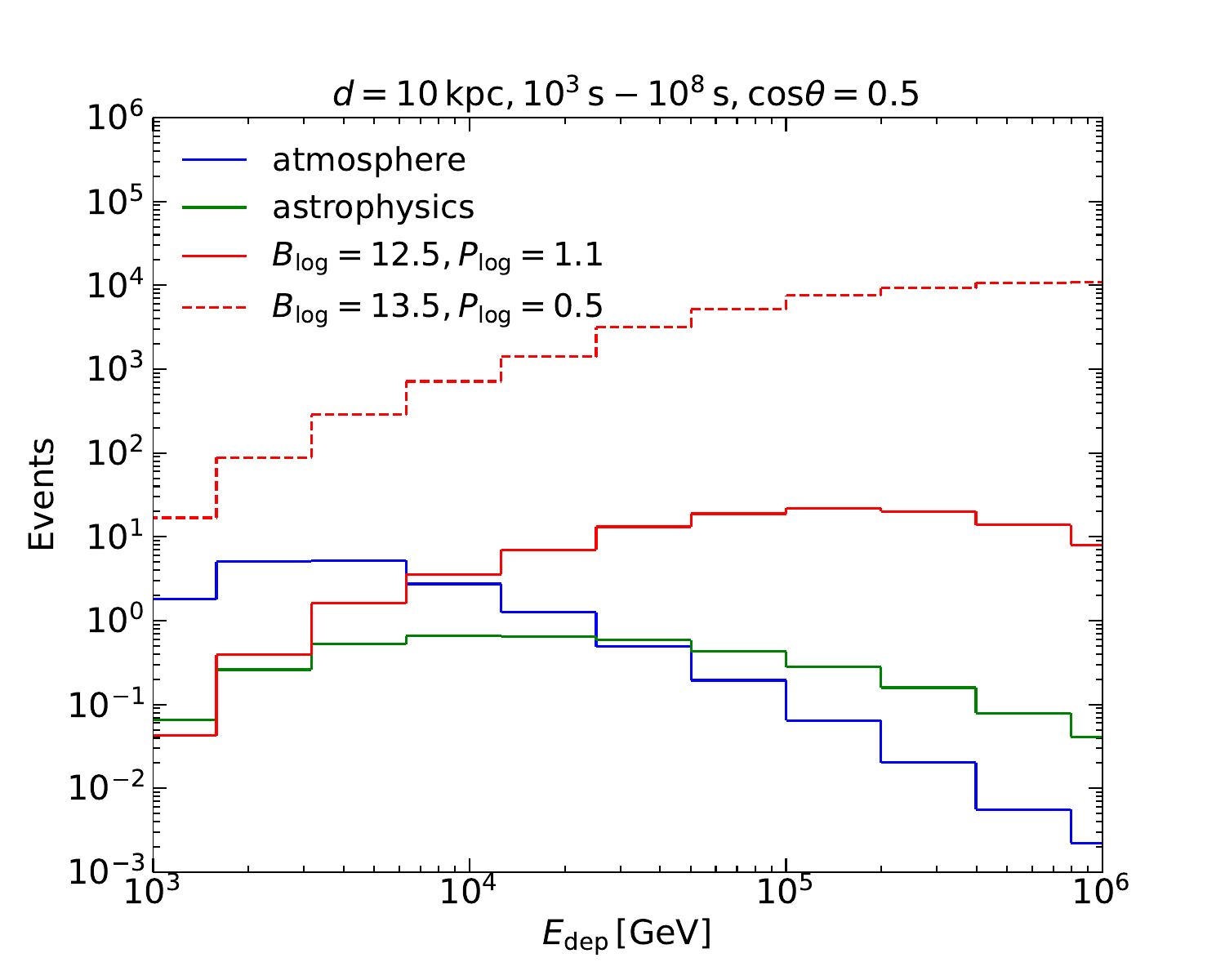}
\includegraphics[width=0.49\textwidth]{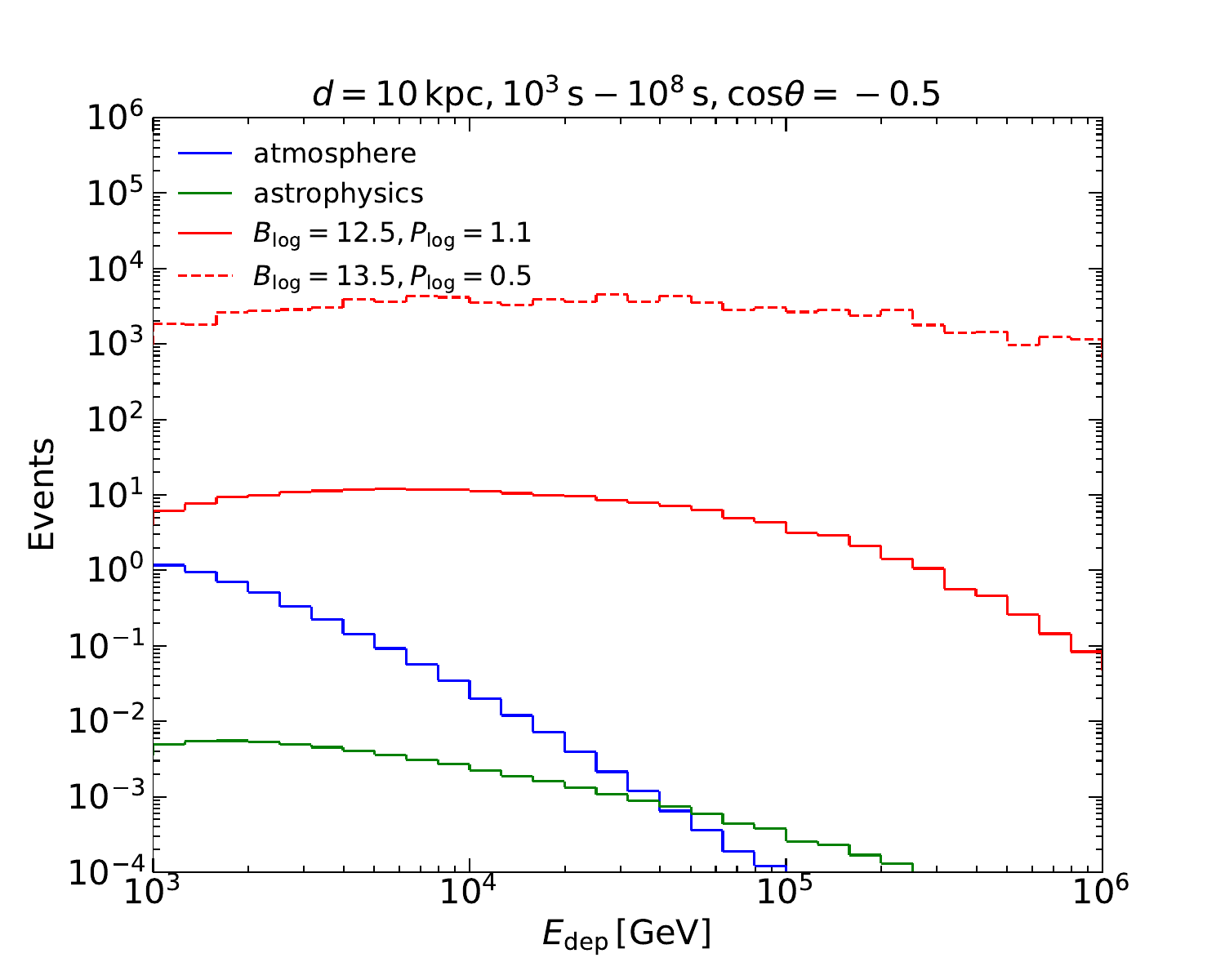}
\caption{Expected spectra of neutrino-induced events at IceCube assuming two pulsar models: model A with $(B_{\rm log}, P_{\rm log}) = (12.5, 1.1)$ and model B with $(B_{\rm log}, P_{\rm log}) = (13.5, 0.5)$. Both downgoing cascades ($\cos\theta=0.5$, left) and upgoging tracks ($\cos\theta=-0.5$, right) are considered.}
\label{fig:eventspec}
\end{figure*}

Figure~\ref{fig:eventspec} presents the expected spectra of downgoing cascades ($\cos\theta=0.5$, left panel) and upgoing tracks ($\cos\theta=-0.5$, right panel) at IceCube, induced by HE neutrinos from a pulsar at 10 kpc. Two pulsar models used in Fig.~\ref{fig:fluence} are considered: model A with $(B_{\rm log}, P_{\rm log })=(12.5, 1.1)$ and model B with $(B_{\rm log}, P_{\rm log })=(13.5, 0.5)$. Due to a higher effective area, the event number for upgoing tracks at low energies is larger than that for cascades. As energy increases, the earth absorption becomes more significant, leading to a suppression in the number of upgoing tracks. As shown in the figure, atmospheric neutrinos dominate at lower energies, while the diffuse astrophysical neutrinos become 
more important above 10 TeV. For pulsar model B, the event spectra exceed the background across the entire energy range explored, which indicate a high detection possibility. With a harder spectrum (spectral index $\sim -1$ up to PeV for model A and up to 100 PeV for model B) compared to those of the atmospheric and diffuse astrophysical neutrinos, the dominance of the signal over background become increasingly significant at higher energies.

The likelihood function is constructed assuming Poisson statistics in each time-energy bin, and is defined as:
\begin{align}
\mathcal{L}(\{s_{ij}\}) = \max_{\{f_b\}} \Bigg\{
&\prod_{{i,j}} {\rm Poiss}  \Big(N_\text{ij}\Big|s_{ij} 
+ (1+f_b) b_{ij}\Big)
\nonumber \\
& \times \exp\Big[-{f_b^2 \over 2\sigma_b^2}\Big] \Bigg\}.
\label{eq:likeli}
\end{align}
Here, ${\rm Poiss}(N\big|\lambda)\equiv \lambda^N e^{-\lambda}/N!$ is the Poisson probability of observing $N$ events given an expected count $\mu$, $N_{ij}$ is the total number of simulated events in time-energy bin $(i,j)$, $s_{ij}$ and $b_{ij}$ are the expected numbers of signals and background in bin $(i, j)$.
In the minimal pulsar scenario, the signal component $s_{ij}=s_{ij}(B,P,d)$ depends on the pulsar's initial period $P$, magnetic field strength $B$, and distance $d$. Under the background-only hypothesis, we set $s_{ij}=0$. To account for the uncertainties in the background modeling, 
we introduce a global nuisance parameter $f_b$ for all bins, representing the uncertainty in the overall background normalization. This parameter is constrained by a Gaussian prior with variance $\sigma_b^2$. For each hypothesis, the likelihood function in Eq.~\eqref{eq:likeli} is always maximized with respect to $f_b$, as indicated. We adopt $\sigma_b=0.1$, noting that our results are largely insensitive to this choice due to the low overall background levels.

To evaluate the discovery potential for HE neutrinos from a newborn pulsar with given $B$, $P$, and $d$, we compute the median discovery significance using the Asimov dataset \cite{Cowan:2010js}. In this approach, the observed counts $N_{ij}$ in Eq.~\eqref{eq:likeli} are set equal to the expected values under the signal-plus-background hypothesis:
$N_{ij} = s_{ij}(B, P, d) + b_{ij}$. 
The test statistic (TS) is then defined as the logarithmic likelihood ratio between the signal-plus-background and the background-only hypotheses:
\begin{align}
\text{TS}(B, P, d) = 2 \Big[&\ln \mathcal{L}\big(\{s_{ij}(B,P,L)\}\big)-\ln \mathcal{L}\big(\{s_{ij}=0\}\big)\Big],
\label{eq:TS-1}
\end{align}
where $\mathcal{L}$ is the likelihood function defined in Eq.~\eqref{eq:likeli}. The median discovery significance for HE neutrinos from the pulsar is given by the square root of the TS value \cite{Cowan:2010js}.

\begin{figure}[htbp]
\centering    \includegraphics[width=0.49\textwidth]{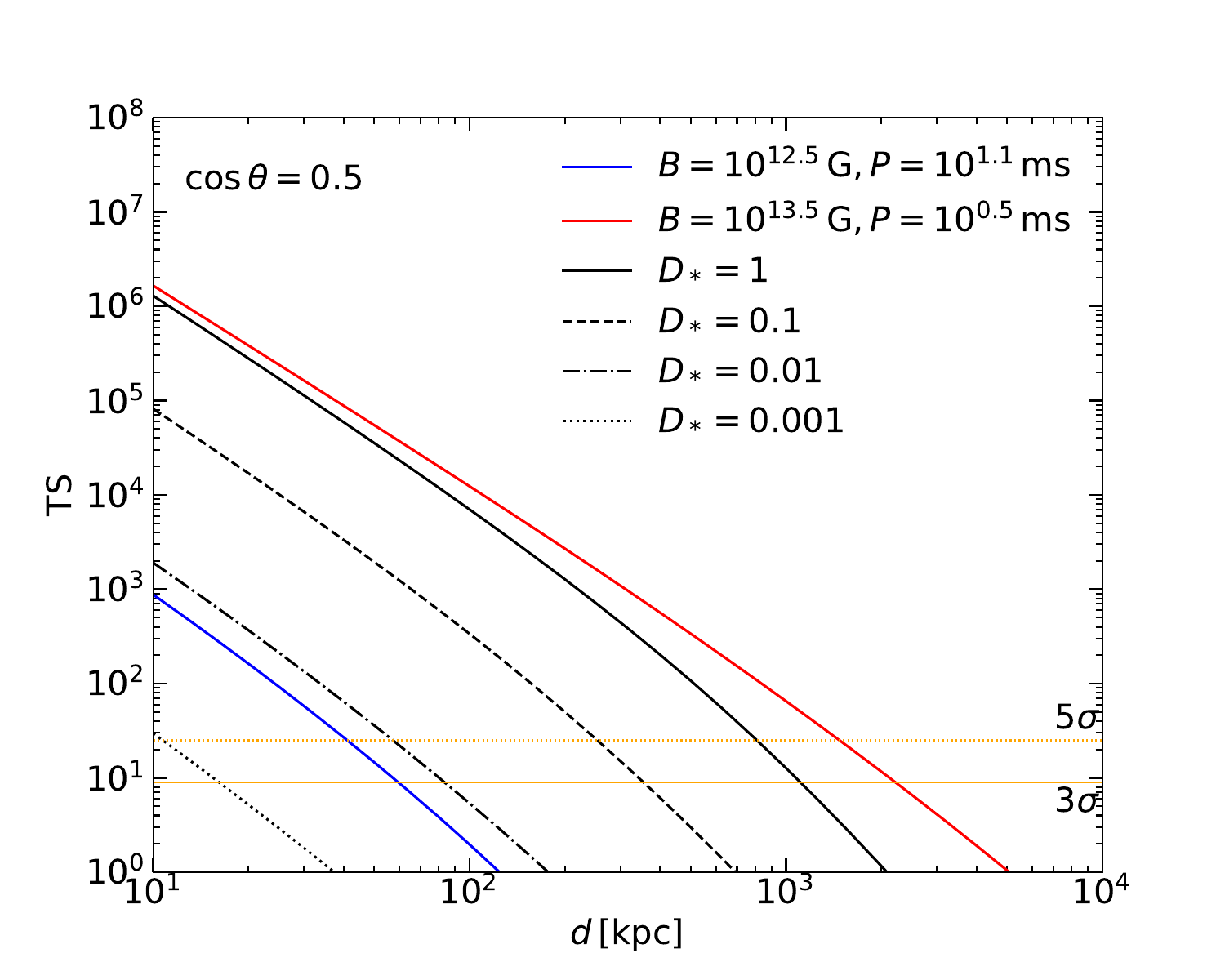}
\caption{Discovery significance of HE neutrinos from newborn pulsars as a function of source distance for pulsar models A (blue) and B (red) based on cascade events. For comparison, the results for the SN ejecta-CSM interaction model, discussed in Sec.~\ref{sec:pulsar-or-not}, are also shown.}
\label{fig:detection-1}
\end{figure}

The test statistic defined in Eq.~\eqref{eq:TS-1}, computed using cascade events, is shown in Fig.~\ref{fig:detection-1} for pulsar models A (blue) and B (red) as a function of source distance $d$. The test statistic decreases with $d$, as the expected signal event number varies with $d^{-2}$. In the regime of large signal statistics (i.e., $\mu_{ij} \gg b_{ij}$ in at least one bin), the test statistic approximately follows $(\mu_{ij}/b_{ij})\ln(\mu_{ij}/b_{ij})$. For typical ratios $\mu_{ij}/b_{ij} \sim 10^{2-6}$, this expression scales roughly as $\mu_{ij}^{1.2-1.3}$, implying an overall distance dependence of approximately $d^{-2.4}$ to $d^{-2.6}$. The horizontal lines indicate discovery significance of $3\sigma$ (${\rm TS}=9$) and $5\sigma$ (${\rm TS}=25$). As shown, the detection horizons for pulsar models A and B reach up to approximately 61 kpc (42 kpc) and 2.3 Mpc (1.5 Mpc) at the $3\sigma$ ($5\sigma$) confidence levels (CLs), respectively.

\begin{figure*}[htbp]
\centering   \includegraphics[width=0.49\textwidth]{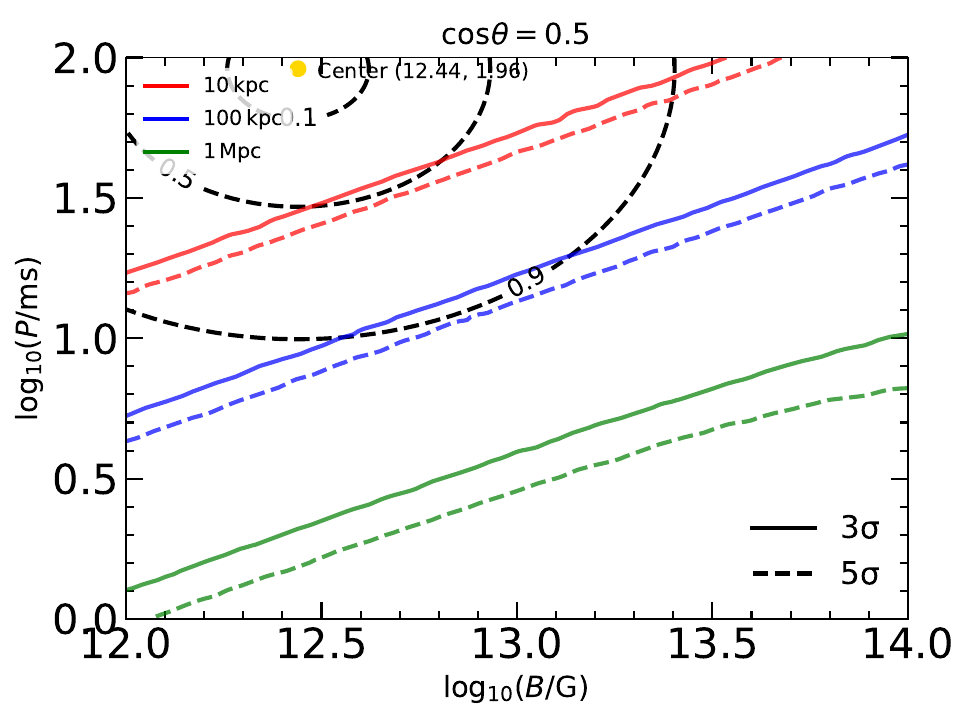}
\includegraphics[width=0.49\textwidth]{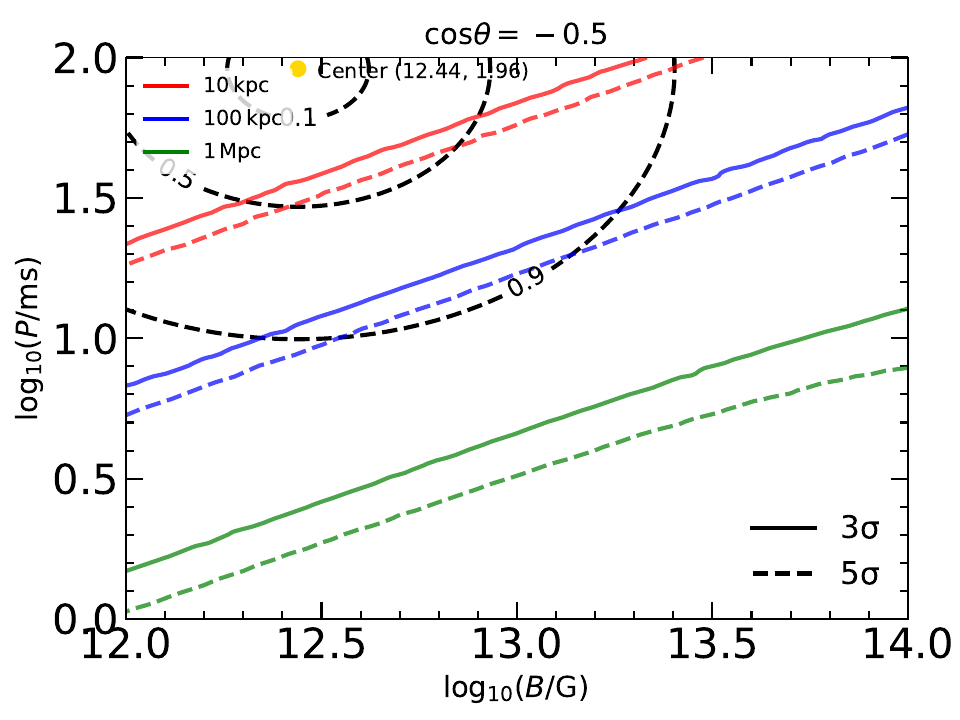} 
\caption{Contours of median discovery significance (3$\sigma$ and 5$\sigma$) for HE neutrinos from a nearby newborn pulsar located at $d=10$ kpc, 100 kpc, and 1 Mpc. For pulsars located at the southern (left) and northern (right) hemispheres, starting cascade events and tracks are used, respectively. The black dashed lines  indicate contours of cumulative probability (0.1, 0.5, and 0.9) for the pulsar population, assuming a binormal distribution in $B_{\rm log}$ and $P_{\rm log}$ [Eq.~\eqref{eq:f_BP}].}
\label{fig:detection}
\end{figure*}

Figure~\ref{fig:detection} shows the contours of median discovery significance at the 3$\sigma$ and 5$\sigma$ CLs in the $B_{\rm log}$-$P_{\rm log}$ parameter space. The calculations are performed for source distances of $d=10$~kpc, 100 kpc, and 1 Mpc, assuming locations in the southern   ($\cos\theta=0.5$, left panel) and northern ($\cos\theta=-0.5$, right panel) hemispheres. As $d$ increases, the detection of HE neutrinos above the background requires higher $B$ and smaller $P$. Specifically, when the spindown time $t_{\rm sd} \gtrsim 10^7$~s,
the expected number of neutrino signals scales with the factor $B/P^2$ (or $B_{\rm log}-2P_{\rm log}$), leading to significance contours that follow constant values of this ratio. For a nearby newborn pulsar at $d=10$ and 100 kpc, the detection of HE neutrinos at the 3$\sigma$ (or 5$\sigma$) CL requires $B_{12}/P^2 \gtrsim 0.003$ (or 0.005) and $\gtrsim 0.034$ (or 0.052), respectively. In contrast, for more distant sources such as those at $d=1$ Mpc, the required high $B$ and low $P$ lead to $t_{\rm sd}\lesssim 10^7$~s, which breaks the $B/P^2$ degeneracy---a behavior clearly visible in the figure. It should be noted that pulsars in the southern sky observed through cascade events provide similar, but slightly higher, detection significance compared to those in the northern sky.
In addition, 
the $3\sigma$ discovery contour shown in the left panel of Fig.~\ref{fig:detection} aligns with the contour of $N_{\rm tot} \sim 5$ cascades, as indicated in Fig.~\ref{fig:Ntot}.

To assess the prospect of detecting HE neutrinos from a future galactic pulsar, it is important to compare the discovery significance contours with the distribution of pulsars in $B$ and $P$. The magnetic field strengths and spin periods of newborn pulsars can be well described by independent lognormal distributions:
\begin{align}
f(B_{\rm log}, P_{\rm log}) = & {1\over \sqrt{2\pi} \sigma_B} e^{-(B_{\rm log}-\langle B_{\rm log} \rangle)^2\over 2\sigma_B^2} \nonumber \\
& \times {1\over \sqrt{2\pi} \sigma_P} e^{-{(P_{\rm log}-\langle P_{\rm log} \rangle)^2 \over 2\sigma_P^2}},\label{eq:f_BP} 
\end{align}
where the mean magnetic field 
and spin periods
are approximately $\langle B_{\rm log} \rangle=12.44$, $\langle P_{\rm log} \rangle=1.96$, with standard deviations $\sigma_B=0.44$ and $\sigma_P=0.53$ \cite{Igoshev_2022}.
The contours of cumulative probability in the $B_{\rm log}$-$P_{\rm log}$ space at 0.1, 0.5, and 0.9 are shown in Fig.~\ref{fig:detection} (black dashed lines), with the mean value indicated by a yellow filled circle. At a distance of $d=10$ kpc, roughly 10\% of the pulsar population lies below the detection contours in the parameter space, which can yield detectable HE neutrino signals. In an optimistic case, with a pulsar located at $d=1$ kpc, the probability of detecting HE neutrinos with IceCube increases to roughly 40\%.

\subsection{diffuse neutrino flux}

The diffuse flux (in unit of $\rm GeV^{-1}~cm^{-2}~s^{-1}~sr^{-1}$) from all newborn pulsars can be expressed as
\begin{align}
\label{eq:diffuse}
\Phi_\nu(E_\nu)=\frac{cf_s}{4\pi}\int_0^{z_{\rm max}} & \mathcal{R}(z)\Big\langle {dN_\nu \over dE_\nu}[E_\nu(1+z)] \Big\rangle \nonumber \\
& \times 
(1+z)\left|\frac{dt}{dz} \right| dz,
\end{align}
where $f_s \sim 0.1$--1 is introduced to accounts for the fraction of pulsars contributing to HE neutrino production and/or the overall uncertainties in the CR normalization,  
and $\langle dN_\nu(E_\nu)/dE_\nu \rangle$ denotes
the neutrino fluence per pulsar, averaged over the distributions of $B$ and $P$:
\begin{align}
\Big\langle {dN_\nu \over dE_\nu}(E_\nu) \Big\rangle=\int\int & {dN_\nu\over dE_\nu}(E_\nu, B, P) \nonumber \\
& \times f(B_{\rm log}, P_{\rm log}) dB_{\rm log} dP_{\rm log}.
\end{align}
The redshift-time relation is encoded in
\begin{align}
\label{eq:dt/dz}
{\left|\frac{dt}{dz}\right| = \frac{1}{H_0(1+z)\sqrt{\Omega_M(1+z)^3+\Omega_\Lambda}},}
\end{align}
with $H_0=67$ km s$^{-1}$ Mpc$^{-1}$, $\Omega_M=0.3$, and $\Omega_\Lambda=0.7$. 

Assuming the pulsar birth rate traces the cosmic star formation history, the pulsar rate $\mathcal{R}(z)$ can be  parameterized as \cite{Madau_2014} 
\begin{align}
\label{eq:local_birth_rate}
\mathcal{R}(z)&=\mathcal{R}_{\rm SFR}(z) \nonumber \\
&=\mathcal{R}(0)(1+z)^{2.7}
\frac{1+[1/2.9]^{5.6}}{1+[(1+z)/2.9]^{5.6}},
\end{align}
where $\mathcal{R}(0) \approx 1.2 \times 10^{-4}~ \text{yr}^{-1}~ \text{Mpc}^{-3}$ is the local pulsar birth rate. Note that 
this value is comparable to the observed rate of CCSNe, a significant fraction of which are expected to result in NS formation~\cite{Heger}. For comparison, we also consider a simplified scenario where the pulsar birth rate is independent on redshift with
$\mathcal{R}(z)=\mathcal{R}_{\rm uniform}(z)=\mathcal{R}(0)$.

\begin{figure}[htbp]
\centering    \includegraphics[width=0.49\textwidth]{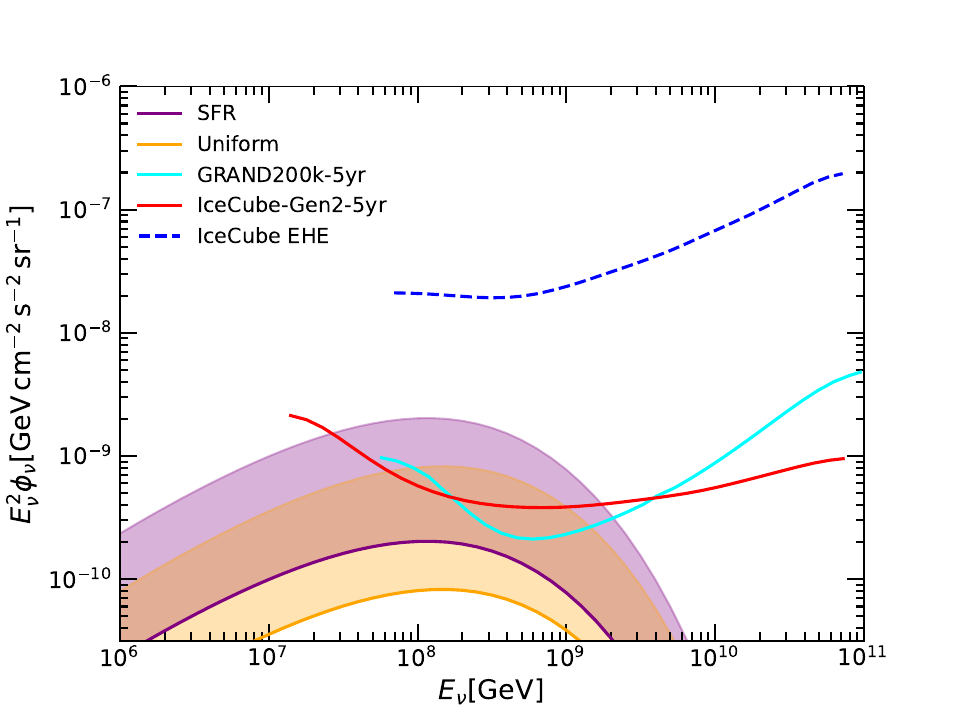}
\caption{The all-flavor diffuse neutrino flux from newborn pulsars. The 5-year sensitivity of IceCube-Gen2~\cite{IC_sensitivity} and the 10-year sensitivity of GRAND200k~\cite{GRAND200k_sensitivity} are indicated by the red and cyan lines, respectively. The blue dashed curve denotes the 90\% CL upper limit on the EHE diffuse flux from nine years of IceCube data~\cite{Aartsen_2018}. Shaded bands correspond to $f_s \in [0.1, 1]$, representing uncertainties in the source contribution, with the purple and orange regions denoting the SFR-traced and uniform birth rate models, respectively.}
\label{fig:diffuse}
\end{figure}

Figure~\ref{fig:diffuse} shows the resulting diffuse HE neutrino flux from newborn pulsars. The shaded purple and orange bands correspond to the results using $\mathcal{R}_{\rm SFR}$ and $\mathcal{R}_{\rm uniform}$, respectively, with the range reflecting $f_s=0.1$ (lower edge) to 1 (upper edge). Both predictions lie below the 90\% CL upper limit from IceCube (blue dashed curve)~\cite{Aartsen_2018}, indicating that current data do not significantly constrain this class of sources \cite{Fang:2014qva}. A substantial portion of the predicted fluxes in the $10^8$--$10^9$ GeV range, however, lies within the projected sensitivities of 
IceCube-Gen2 (5-year)~\cite{IC_sensitivity} and GRAND200k (10-year)~\cite{GRAND200k_sensitivity}, shown by the red and cyan curves.
Future observations will thus be critical for probing the pulsar contribution to the diffuse HE neutrino flux. A continued non-detection over the next decade would begin to disfavor the most optimistic scenarios ($f_s\rightarrow1$).

\begin{figure*}[htbp]
\centering   \includegraphics[width=0.49\textwidth]{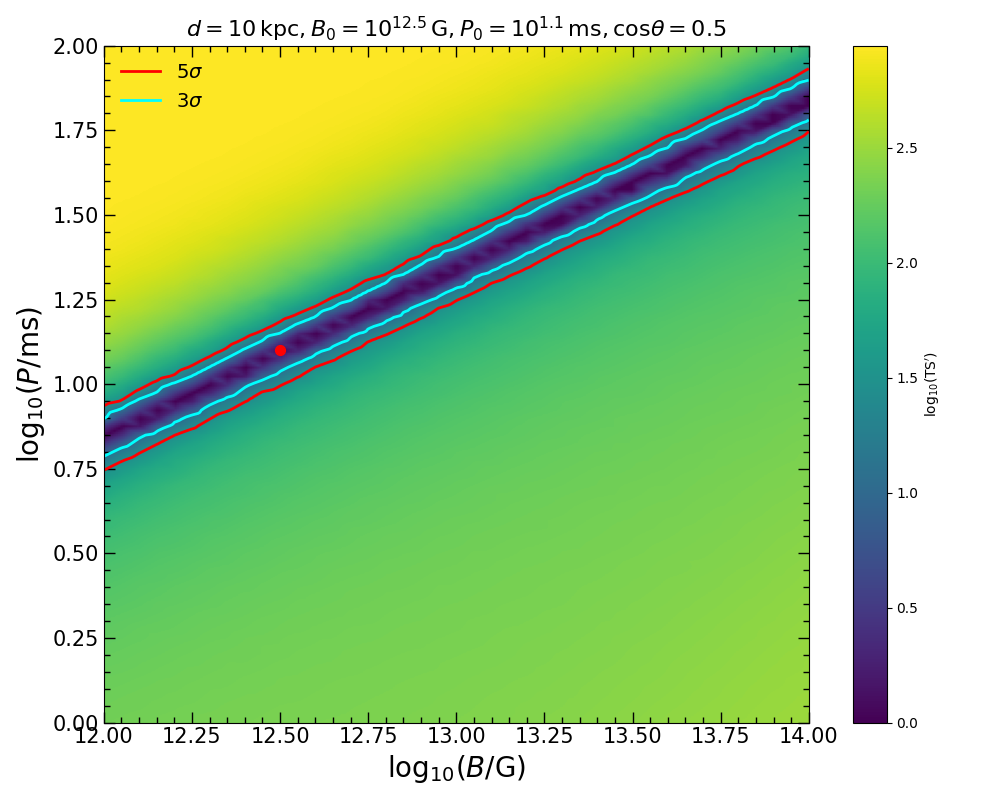}
\includegraphics[width=0.49\textwidth]{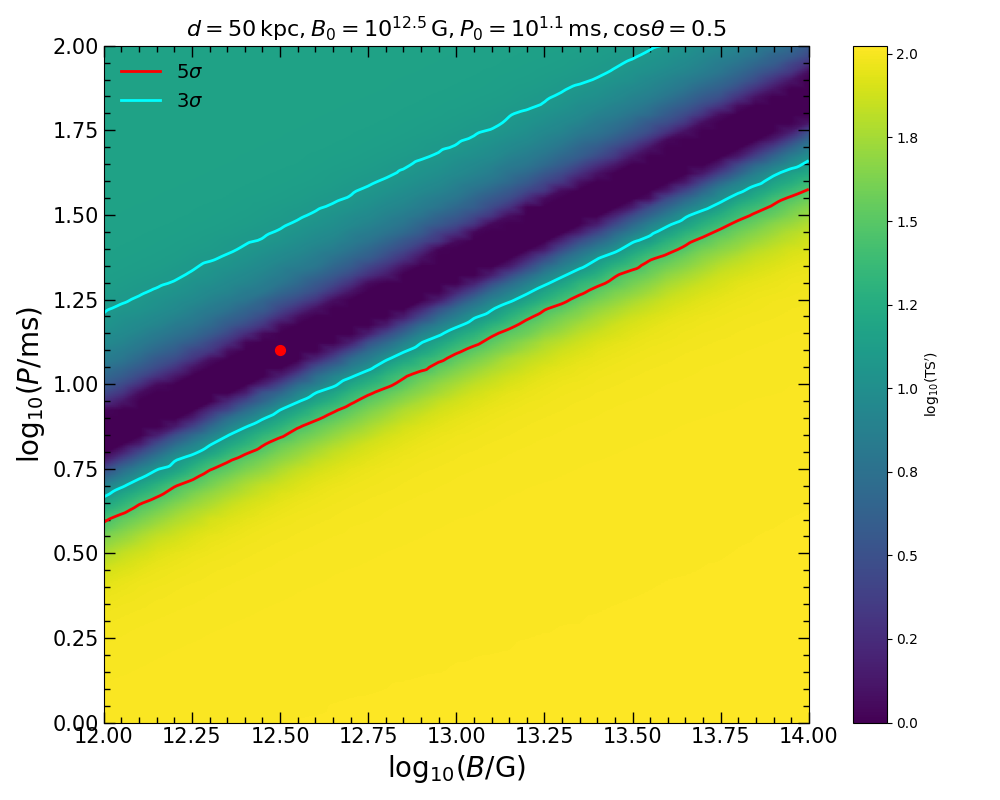} \\
\includegraphics[width=0.49\textwidth]{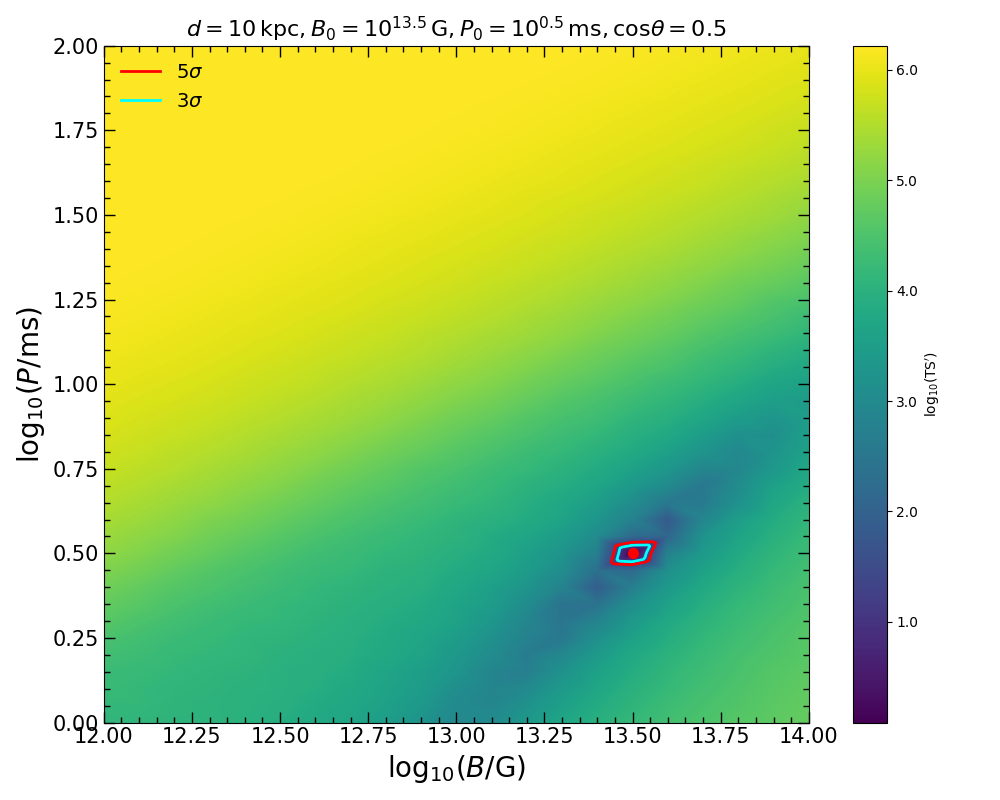}
\includegraphics[width=0.49\textwidth]{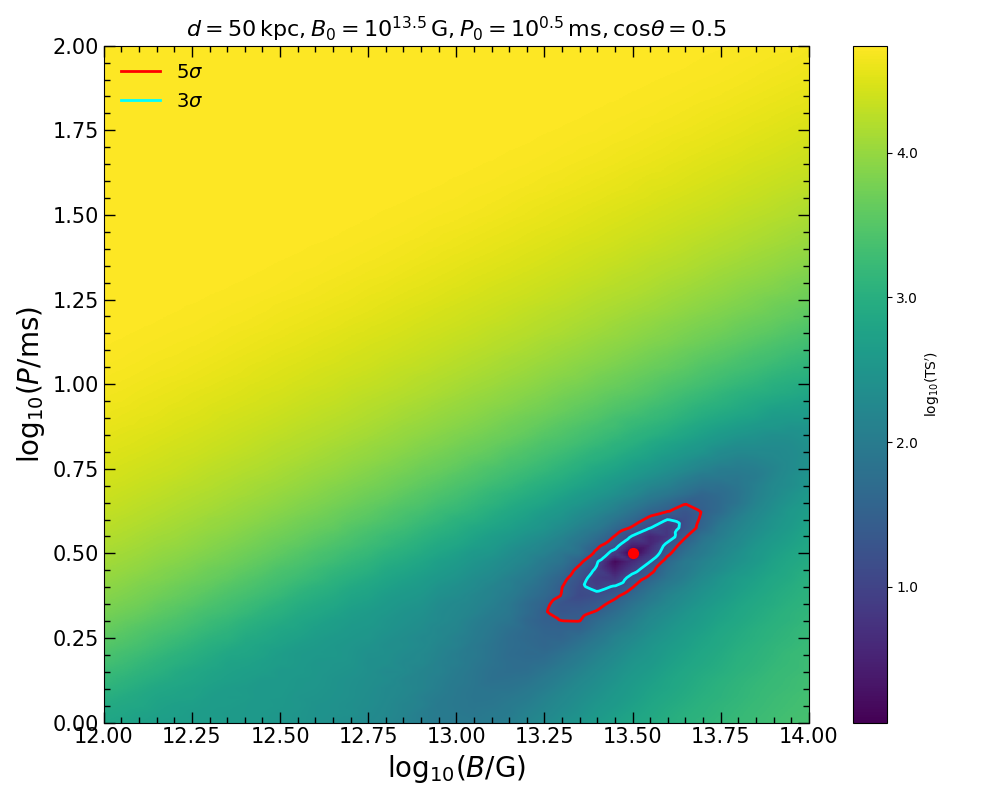}
\caption{The expected $3\sigma$ and $5\sigma$ significance contours for measuring the magnetic
field strength $B$ and initial spin period $P$ assuming pulsar models A (upper) and B (lower), marked by red dots. Pulsars are assumed to located in the southern hemisphere with $\cos\theta=0.5$, and downgoing cascades are considered.}
\label{fig:telling-pulsar}
\end{figure*}

\section{probing the magnetic field and initial period}
\label{sec:pulsar properties}

Following the likelihood approach constructed in Sec.~\ref{sec:signals}, we assess the potential to measure a pulsar's magnetic field and initial spin period from its neutrino signal in this section.

Introducing a nuisance parameter $f_s$ to account for the overall uncertainty in modeling the predicted HE neutrino flux, the likelihood function introduced in Eq.~\eqref{eq:likeli} is extended as:
\begin{align}
\mathcal{L}'(\{s_{ij}\}) = \min_{\{f_s, f_b\}} \Bigg\{
&\prod_{{i,j}} {\rm Poiss}  \Big(N_\text{ij}\Big|(1+f_s)s_{ij}+(1+f_b)b_{ij}\Big)
\nonumber \\
& \times \exp\Big[-{f_s^2 \over 2\sigma_s^2}-{f_b^2 \over 2\sigma_b^2}\Big] \Bigg\},
\label{eq:likelip}
\end{align}
where $\sigma_s$ is the standard deviation for the signal uncertainty, and the maximization is performed with respect to both $f_s$ and $f_b$. We set $\sigma_s=0.1$ and $\sigma_b=0.1$, and note that $\sigma_b$ has only a minor impact on the results.

As in Sec.~\ref{sec:signals}, the event number $N_{ij}$ is set to $s_{ij}(B_0,P_0,d)+b_{ij}$ for a pulsar characterized by benchmark parameters $B_0$, $P_0$, and $d$. To quantify how precisely the magnetic field strength and initial spin period can be constrained,\footnote{The distance to the nearby pulsar is assumed to be measured through optical observations, with its associated uncertainty incorporated into the likelihood analysis via the nuisance parameter $f_s$.} we compare the likelihood function under two hypothesis with two different parameter sets: one corresponding to the benchmark parameter set ($B_0$, $P_0$), and the other to a test set ($B$, $P$). The corresponding test statistic is given by: 
\begin{align}
{\rm TS}'(B_0, P_0, B, P, d) =& 2 \Big[\ln{\cal L}'\big(\{s_{ij}(B_0, P_0, d)\}\big) \nonumber \\
& -\ln{\cal L}'\big(\{s_{ij}(B, P, d)\}\big)\Big].
\end{align}
With $B$ and $P$ treated as free parameters, a 3$\sigma$ and  5$\sigma$ deviation correspond to test statistic values of approximately 11.83 and 28.74, respectively.

Figure~\ref{fig:telling-pulsar} shows the expected $3\sigma$ and $5\sigma$ confidence contours in the $B_{\rm log}$--$P_{\rm log}$ plane for constraining the magnetic field strength and initial spin period based on the HE neutrino signals. Results are shown for pulsar models A (top) and B (bottom), at distances of 10 kpc (left) and 50 kpc (right). The source is assumed to be located in the southern hemisphere ($\cos\theta=0.5$), with starting cascade events used in this analysis. Similar result is expected for sources in the northern hemisphere using upgoing tracks, as indicated in Fig.~\ref{fig:detection}.

For model A with relatively lower $B$ and higher $P$, the corresponding $t_{\rm sd}$ is larger than $10^7$~s. As a result, the $3\sigma$ and $5\sigma$ contours exhibit a pronounced degeneracy along lines of constant $B/P^2$, consistent with the scaling behavior of proton injection rate and energy described in Eqs.~\eqref{eq:Ept} and \eqref{eq:Npdot}. For this particular case, the HE neutrino signal primarily constrains the combination $B/P^2$, rather than 
$B$ and $P$ individually.
At a distance of $d=10$ kpc, two distinct $3\sigma$ and $5\sigma$ contours appear on either side of the benchmark parameter point (marked by a red dot). For model A, where $B_{12}/P^2 \approx 0.02$, this combination can be constrained to the range $[0.016, 0.027]$ ($[0.013, 0.032]$) at the $3\sigma$ ($5\sigma$) CL. To achieve a 20\% precision at the same distance, we find that pulsars must have $B_{12}/P^2 \gtrsim 0.08$. As $d$ increases to 50 kpc for model A, the event statistics for HE neutrinos from the pulsar become too low for reliable identification, resulting in only one visible contour on the higher-$B$ side for both the $3\sigma$ and $5\sigma$ levels. In this case, the analysis yields only an upper bound on $B/P^2$, rather than a closed confidence interval.

With a stronger magnetic field $B$ and shorter spin period $P$ for model B, the degeneracy in $B/P^2$ is partially broken due to the nontrivial temporal evolution of the neutrino signal at times
$t \gtrsim t_{\rm sd}$. This leads to closed confidence contours, as shown in the lower panels of Fig.~\ref{fig:telling-pulsar}. At a distance of $d=10$ kpc, the event statistics are sufficiently high to enable both $B$ and $P$ to be constrained within $\sim 10$\% around the assumed benchmark values $B_0$ and $P_0$. At $d=50$ kpc, the reduced statistics lead to broader contours that are elongated along lines of constant $B/P^2$, indicating a residual degeneracy. As shown in Fig.~\ref{fig:telling-pulsar}, a precise determination of $B$ and $P$ from HE neutrino observations requires sufficiently strong magnetic fields and rapid rotation. 
To measure both $B$ and $P$ with better than 20\% precision at the $3\sigma$ CL for a source at $d=10$ kpc, the ratio
$B_{12}/P^2$ is required to exceed $\sim 0.35$.

\begin{figure*}[htbp]
\centering   \includegraphics[width=0.49\textwidth]{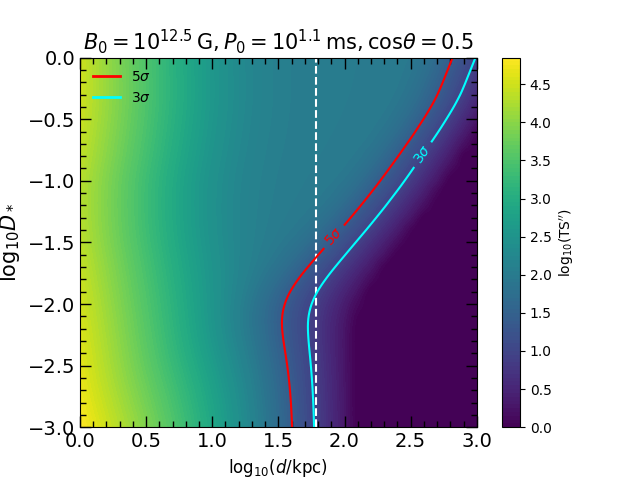}
\includegraphics[width=0.49\textwidth]{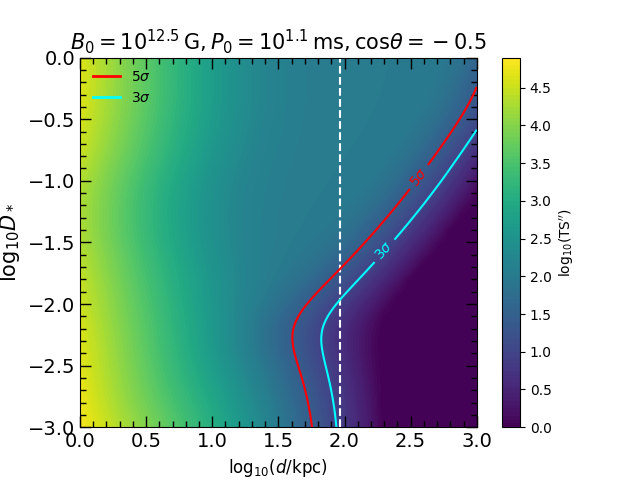} \\
\includegraphics[width=0.49\textwidth]{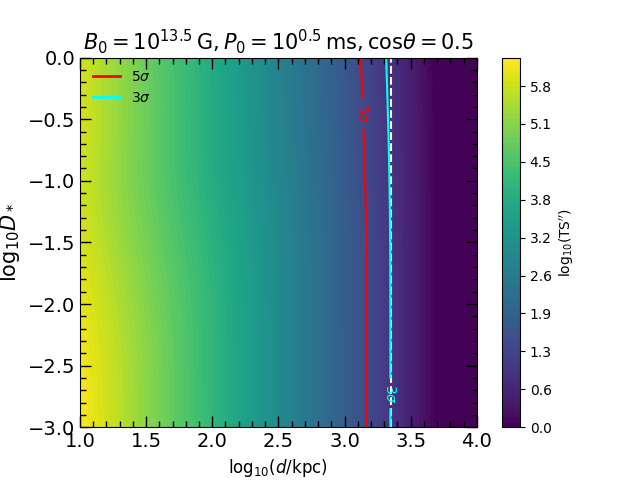}
\includegraphics[width=0.49\textwidth]{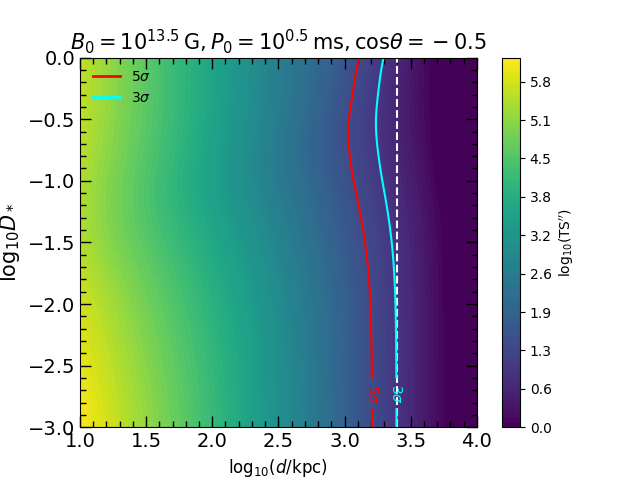}
\caption{The expected median significance of distinguishing the minimal pulsar model from the SN ejecta-CSM interaction model. Pulsars are assumed to located in the southern hemisphere with $\cos\theta=0.5$ and downgoing cascades are considered.}
\label{fig:telling-pulsar-SN}
\end{figure*}

\section{pulsar origin or ejecta-CSM interaction}
\label{sec:pulsar-or-not}

As previously discussed, HE neutrinos are also expected to be produced when protons, accelerated at shocks driven by SN ejecta, undergo hadronic interactions with the surrounding CSM \cite{Murase:2010cu,Murase:2017pfe,Murase:2018okz}. This mechanism operates independently of pulsar formation. For CCSNe embedded in dense CSM, the fluxes and spectra of HE neutrinos could be comparable to those predicted by the pulsar model.
In this section, we investigate in detail the potential to differentiate between these two distinct HE neutrino production mechanisms.   

The flux of HE neutrinos resulting from interactions between SN ejecta and the CSM depends sensitively on the CSM density. Assuming a wind-like profile, the CSM density can be expressed as \cite{Murase:2017pfe}
\begin{align}
\rho_{\rm CSM}(r) = 5 \times 10^{16}~{\rm g~cm^{-1}} D_* r^{-2}.
\end{align}
where $D_*$ is a dimensionless parameter characterizing the CSM density. For SN types Ibc, IIb, and II-P, $D_*$ typically ranges from $10^{-5}$ to $10^{-3}$, whereas for type IIn, it can be as large as $10^{-2}$--1. For simplicity, we use the analytical expressions provided in Eq.~(7) of \cite{Murase:2017pfe} to estimate the HE neutrino fluence. These approximations are shown to agree well with the exact numerical results and are sufficient for the purpose of our study. We adopt standard values for the relevant parameters, such as the total kinetic energy and mass of the SN ejecta ($\varepsilon_{\rm ej}=10^{51}$~erg and 
$M_{\rm ej} = 10M\odot$), and a spectral index $s=2.2$ for accelerated protons. Under these assumptions, the all-flavor neutrino fluence can be expressed in terms of $D_*$ as:
\begin{align}
E_\nu^2F_\nu^{\rm SN} \sim & \, 83 \, \text{GeV cm}^{-2} \min[1, f_{pp}^{\rm SN}] \, \epsilon_{\text{cr}, -1} \left(\frac{E_\nu}{0.4 \, \text{GeV}}\right)^{-0.2} \nonumber \\
& \times D_{*,-2}^{7/10} t_{5.5}^{7/10} \left(\frac{d}{10 \, \text{kpc}}\right)^{-2},
\end{align}
where $f_{pp}^{\rm SN} \simeq 0.82 D_{*,-2}^{6/5}t_{5.5}^{-4/5}$ is the optical depth for HE protons interacting with the CSM, and $\epsilon_{\rm cr} \sim 0.1$ denotes the fraction of kinetic energy converted into CRs by shock dissipation. For reference, the $3\sigma$ detection horizon for the SN interaction model with IceCube spans from $\sim 10$ kpc for $D_*=10^{-3}$ to $\sim 1$ Mpc for $D_*=1$, as shown in Fig.~\ref{fig:detection-1} (see also Ref.~\cite{Valtonen-Mattila:2022nej}).

To investigate the potential for distinguishing between the two HE neutrino production scenarios, we construct an Asimov dataset $\{N_{ij}\}$, assuming neutrino emission from the pulsar model characterized by parameters $(B_0, P_0, d)$, and then evaluate how strongly the pulsar model is favored over the SN interaction model. The corresponding test static is defined as
\begin{align}
{\rm TS}''(B_0, P_0, D_*, d) =& 2 \Big[\ln{\cal L}'\big(\{s_{ij}(B_0, P_0, d)\}\big) \nonumber \\
& -\ln{\cal L}'\big(\{\bar s_{ij}(D_*, d)\}\big)\Big].
\end{align}
where ${\cal L}'$ is the likelihood function defined in Eq.~\eqref{eq:likelip}, and $\bar s_{ij}(D_*, d)$ denotes the expected number of signal events in each time-energy bin under the SN ejecta-CSM interaction model. For given values of $B_0$, $P_0$ and $d$, the SN interaction model is considered disfavored at the $3\sigma$ and $5\sigma$ CLs if ${\rm TS}''$ exceeds 9 and 25, respectively.     

Figure~\ref{fig:telling-pulsar-SN} shows the values of ${\rm TS}''$ for distinguishing between the two models in the $\log_{10}(d/{\rm kpc})$--$\log_{10}D_*$ plane. As in Sec.~\ref{sec:pulsar properties},
the analysis is performed assuming pulsar models A (upper panels) and B (lower panels).
Results are shown for two sky locations—southern hemisphere ($\cos\theta=0.5$, left panels) and the northern hemisphere ($\cos\theta=-0.5$, right panels).
The sky-blue and red contours denote the $3\sigma$ (${\rm TS}''=9$) and $5\sigma$ (${\rm TS}''=25$) CLs for discriminating the pulsar scenario from the SN interaction model. To the left of these contours, the SN interaction model characterized by a given $D_*$ at distance $d$ is disfavored. In contrast, in the right-hand region, where $d$ is large and event statistics are low, the two models can not be effectively distinguished. For reference, the white dashed vertical lines indicate the distance at which the pulsar model with parameters
$(B_0, P_0)$ yields a detectable neutrino signal above background at the $3\sigma$ CL, as shown in Fig.~\ref{fig:detection-1}.

For small values of $D_*$, the expected number of signal events from the SN interaction model is much lower than that from the pulsar scenario. Consequently, the test statistics for favoring the pulsar model over the SN interaction model closely resembles that for favoring it over the background-only hypothesis. This leads to an overlap between the $3\sigma$ detection contour (white lines) and the $3\sigma$ model comparison contour (sky-blue lines), particularly in the case of pulsar model B, which predicts a higher neutrino event rate.    

The expected event count from the interaction model increases with $D_*$. To disfavor this model, higher event statistics, or equivalently, a smaller distance $d$ is required. This trend is reflected in the shift of the $3\sigma$ and $5\sigma$ contours towards smaller $d$ as $D_*$ increases. Once $\log_{10}D_*$ exceeds approximately $-2.2$, the SN interaction model begins to yield more events than the pulsar model A, making it easier to distinguish between them. As a result, the contours shift back toward larger $d$ with increasing $D_*$, as seen in the upper panel of Fig.~\ref{fig:telling-pulsar-SN}. This turnover does not occur for pulsar model B (lower panels), where the pulsar scenario dominates in event count across the full range of $D_*$, maintaining its separation from the interaction model. For both pulsar models A and B, the two production scenarios can be clearly distinguished at a typical galactic distance of $d=10$ kpc. We have also verified that for any pulsar model producing detectable HE neutrino signals, the emission can be clearly distinguished from that of the SN ejecta-CSM interaction model based on their temporal and spectral characteristics.

\section{Discussions and summary}
\label{sec:summary}

In this work, we have studied the production and detection prospects of HE neutrinos originating from newborn pulsars formed in CCSNe. We adopt a minimal pulsar scenario, in which protons accelerated by relativistic pulsar winds interact with the expanding SN ejecta, producing HE neutrinos via $pp$ interactions. These neutrinos typically emerge within weeks to months following the explosion and span a broad energy range from TeV to EeV.

We take IceCube as the reference detector and consider both downgoing cascades and upgoing tracks for pulsars located in the southern and northern hemispheres, respectively. To evaluate the detection prospects, we perform a binned likelihood analysis incorporating both energy and temporal information. For two representative pulsar model: a typical case (model A) with $(\log_{10}(B/{\rm G}), \log_{10}(P/{\rm ms})) \equiv (B_{\rm log}, P_{\rm log })=(12.5, 1.1)$, and an optimistic one (model B) with $(13.5, 0.5)$, we find that the $3\sigma$ ($5\sigma$) detection horizon extends to $\sim 61$ kpc (42 kpc) and $\sim 2.3$ Mpc (1.5 Mpc), respectively. By scanning over $B$ and $P$, we show that pulsars within 10 kpc and satisfying $B_{12}/P^{2} \equiv (B/10^{12} {\rm G})(P/{\rm ms})^{-2}\gtrsim 0.003$, which represents about 10\% of the pulsar population, can produce HE neutrino signals detectable above atmospheric and diffuse astrophysical neutrino backgrounds at $\ge 3\sigma$ CL. 
Next-generation neutrino observatories such as IceCube-Gen2 \cite{IceCube-Gen2:2020qha}, KM3NeT \cite{KM3Net:2016zxf}, and TRIDENT \cite{TRIDENT:2022hql} will feature significantly improved effective areas for detecting HE neutrinos. By simply scaling the IceCube effective areas up by a factor of 5, we find that the detection horizon extends by a factor of $\sim 2$, and the probability of observing HE neutrinos from a galactic pulsar increases to $20\%-30\%$.

We have also estimated the diffuse HE neutrino flux from the cosmological population of newborn pulsars. Under optimistic assumptions, the predicted flux lies within the sensitivity range of IceCube-Gen2 and GRAND200k, particularly in the 0.1--1 EeV energy range. These instruments will therefore play a crucial role in testing and constraining the contribution of newborn pulsars to the diffuse HE neutrino background.

Additionally, we explored the potential of HE neutrino signals to measure the magnetic field strength and initial spin period of newborn pulsars. For pulsar model A ($B_{12} /P^2\approx 0.02$) at a distance of 10 kpc, we find that the combination $B/P^2$ can be reconstructed with an uncertainty of approximately 30\% at the $3\sigma$ CL. In contrast, for the more optimistic model B with stronger magnetic field and faster rotation ($B_{12} /P^2\approx 3.16$), the temporal evolution of the neutrino signal helps break the degeneracy between $B$ and $P$, enabling both parameters to be individually constrained to within approximately 10\%, assuming the same source distance. By varying $B$ and $P$, we find that for galactic pulsars at 10 kpc,  $B_{12}/P^2$ must exceed $\sim 0.08$ to achieve a $20\%$ precision on the measurement of $B/P^2$, and exceed $\sim 0.35$ to constrain $B$ and $P$ individually within 20\%.

Finally, we investigated the ability to distinguish the pulsar scenario from an alternative neutrino production model where HE neutrinos arise from hadronic interactions between SN ejecta and the CSM. Applying the same binned likelihood method, we find that for pulsar models A and B located in the Milky Way, the two production models can be discriminated with $\gtrsim 5\sigma$ significance, owing to their distinct temporal and spectral characteristics. For pulsar model B, where the neutrino yield is much higher, this discrimination capability could extend to distances as far as $\sim 1.4$ Mpc. We further confirm that for any pulsar model producing a detectable HE neutrino signal, the ejecta-CSM interaction scenario can be clearly ruled out.

In summary, our
study demonstrates that current and next-generation neutrino telescopes can not only detect HE neutrinos from newborn pulsars but also constrain key pulsar parameters  such as the magnetic field strength $B$ and initial spin period $P$. Moreover, these signals provide a means to distinguish between different  neutrino production scenarios. These capabilities highlight the unique role of HE neutrinos in advancing our understanding of neutron star formation and the associated particle acceleration processes in these extreme astrophysical environments. Note that several limitations remain in this work and merit further investigation. The minimal pulsar model adopted here assumes a pure proton composition and spherically symmetric outflows, neglecting possible contributions from heavy nuclei and effects arising from anisotropic emission. Additionally, more realistic detector modeling--including energy and angular resolution, background rejection, and flavor identification--should be incorporated in future studies to better quantify the detection prospects and the sensitivity to the relevant physical parameters.

\begin{acknowledgments}
This work was supported in part by the National Natural Science Foundation of China (No.~12205258), Guangdong Basic and Applied Basic Research Foundation (No.~2025A1515011082), the ``CUG Scholar" Scientific Research Funds at China University of Geosciences (Wuhan) [No.~2021108].
\end{acknowledgments}

%

\end{document}